\documentclass[preprint,review,14pt]{elsarticle}
\usepackage{lineno}
\usepackage{graphicx}
\usepackage{epsfig}
\usepackage{amsmath,amsfonts, amsthm, mathrsfs, amssymb, bbm}
\usepackage{natbib}
\usepackage{times}
\usepackage{subfigure}
\journal{Journal of Theoretical Biology}

\begin{document}

\begin{frontmatter}

\title{Evolutionary Kuramoto Dynamics}

\author[DartMath]{Elizabeth A. Tripp}
\ead{Elizabeth.A.Tripp.GR@dartmouth.edu}
\author[DartMath,DartBMDS]{Feng Fu}
\ead{fufeng@gmail.com}
\author[DartMath]{Scott D. Pauls\corref{sdp}}
\ead{Scott.D.Pauls@dartmouth.edu}

\address[DartMath]{Department of Mathematics, Dartmouth College, Hanover, NH 03755, USA}

\address[DartBMDS]{Department of Biomedical Data Science, Geisel School of Medicine at Dartmouth, Lebanon, NH 03756, USA}

\cortext[sdp]{Corresponding author at: Professor Scott Pauls, 27 N. Main Street, 6188 Kemeny Hall, Department of Mathematics, Dartmouth College, Hanover, NH 03755, USA. Tel: +1 (603) 646 1047, Fax: +1 (603) 646 1312}

\begin{abstract}
Common models of synchronizable oscillatory systems consist of a collection of coupled oscillators governed by a collection of differential equations.  The ubiquitous Kuramoto models rely on an {\em a priori} fixed connectivity pattern facilitates mutual communication and influence between oscillators.   In biological synchronizable systems, like the mammalian suprachaismatic nucleus, enabling communication comes at a cost --- the organism expends energy creating and maintaining the system --- linking their development to evolutionary selection.  Here, we introduce and analyze a new evolutionary game theoretic framework modeling the behavior and evolution of systems of coupled oscillators. Each oscillator in our model is characterized by a pair of dynamic behavioral traits: an oscillatory phase and whether they connect and communicate to other oscillators or not. Evolution of the system occurs along these dimensions, allowing oscillators to change their phases and/or their communication strategies.  We measure success of mutations by comparing the benefit of phase synchronization to the organism balanced against the cost of creating and maintaining connections between the oscillators. Despite such a simple setup, this system exhibits a wealth of nontrivial behaviors, mimicking different classical games -- the Prisoner's Dilemma, the snowdrift game, and coordination games -- as the landscape of the oscillators changes over time.  Despite such complexity, we find a surprisingly simple characterization of synchronization through connectivity and communication:  if the benefit of synchronization $B(0)$ is greater than twice the cost $c$, $B(0) > 2c$, the organism will evolve towards complete communication and phase synchronization.  Taken together, our model demonstrates possible evolutionary constraints on both the existence of a synchronized oscillatory system and its overall connectivity.

\end{abstract}

\begin{keyword}
 Evolutionary game theory \sep Neuroscience \sep Kuramoto dilemma \sep Cooperation \sep Synchronization \end{keyword}

\end{frontmatter}


\section{Introduction}
\label{intro}

The mammalian suprachiasmatic nucleus (SCN) is a small center in the brain that sits just above the optic chiasm. It receives light/dark signals from the optic nerve and uses them to generate and maintain the organism's circadian rhythm. Most of the roughly 20,000 neurons in the SCN are oscillatory, exhibiting approximately 24--hour rhythms. When their connectivity is disrupted, the neurons oscillate with about the same period but randomly out of phase. However, when connectivity is intact, the oscillations exhibit phase--locked synchronization \cite{welsh1995}. A coherent circadian signal within an organism confers many advantages, as it allows prediction of the light/dark cycle. Mammals can anticipate changes in light that allow them to avoid predators, find food, and generally increase their chances of survival. This basic principle informs natural selection, but its impact on the structure of the SCN is unknown. Mechanisms to generate circadian signals exist in a wide range of species -- including mammals \cite{panda2002circadian}, fruit flies \cite{panda2002circadian}, and cyanobacteria \cite{clodong2007functioning} -- demonstrating that for many, the benefit of such a system outweighs the evolutionary cost. However, for other organisms, like the eyeless Mexican cavefish, developing a circadian clock would not confer the same kind of benefit, and thus such a system did not evolve \cite{moran2014eyeless}.

Mathematically, the mammalian SCN can be viewed as a network of coupled oscillators. Understanding the synchronization of systems of coupled oscillators has a rich history in the study of dynamical systems and applications in numerous fields~\cite{strogatz1kuramoto,arenas2008synchronization}. Decades of work has demonstrated the interplay between the properties of these systems and their ability to synchronize (surveys of the field include \cite{dorfler2014,rodrigues2016,tang2014synchronization}). One of the simplest and most fruitful modeling approaches uses a differential equations system model first introduced by Kuramoto \cite{kuramoto1975}. If $\phi_j$ denotes the oscillatory parameter of neuron $j$, and $\nu_j$ is its intrinsic frequency, the model reads:

\[\dot{\phi_j} = \nu_j + \sum_{k} h_{jk} \sin(\phi_k-\phi_j),\]

\noindent where $h_{jk}$ is the coupling weight between neurons $j$ and $k$. The effect of the coupling term is to pull the oscillations of connected neurons towards one another. 

While a simple and elegant analytic approach exists in the case of two oscillators, with more oscillators and more complicated connectivity, the problem becomes (much) harder. While we can understand this system (and some variants) analytically when the coupling topologies are particularly simple \cite{rodrigues2016} and/or when we look at the mean--field limit as the number of oscillators tends to infinity \cite{rodrigues2016,dorfler2014}, more complex (and biologically plausible) connectivity patterns are not.  For these cases, we must rely on numerical approximation of solutions which can be both difficult and costly computationally.

In this paper we approach the problem using techniques from the field of evolutionary game theory (EGT), which applies classical game theory to the study of evolving populations~\cite{smith1982evolution,hofbauer1998evolutionary,evol_dynamics}. The competitive advantages of various traits (or \textit{strategies}, in the language of EGT) that exist in the population are based on payoffs accrued from pairwise or multi-person game interactions between connected individuals. Traits of individuals are allowed to evolve over time, mimicking the biological process of natural selection. In this way, traits that confer an individual a competitive advantage have greater success at propagating as the population evolves. This framework is designed to answer questions about which population traits are more evolutionarily successful, and under what conditions particular traits are advantageous. When we view the neurons of the SCN as our population of interest, this EGT setup lends itself naturally to our main question: what conditions allow for the synchronization behavior observed in the mammalian SCN to arise? The benefit conferred to the organism from this synchronization is shared by each neuron. However, the communication between neurons that is necessary to ensure this synchronization behavior is costly. Thus, each neuron faces the choice of whether or not to expend the necessary effort to communicate and aid the synchronization effort of the SCN. This kind of trade-off is captured in classic cooperative dilemmas, such as the Prisoner's Dilemma game~\cite{evol_dynamics,axelrodt2006evolution}.

Antonioni and Cardillo \cite{EKD} incorporate EGT into the Kuramoto framework using a modified version of the Prisoner's Dilemma. In their model, neurons have two possible strategies:  cooperation, where neurons influence each other to move towards synchronization, and defection, where they don't.   They use solutions of Kuramoto coupled oscillator systems to calculate payoffs of a neuron's strategy based the level of local synchronization.  In contrast to the typical EGT setup, where an individual's payoff is dependent on the current state of the system and the strategy of their opponent, in this framework,  a neuron will receive the same payoff regardless of the strategy of its particular opponent as the payoff function depends on the the strategies and phases of all of its neighbors.

While Antonioni and Cardillo's framework captures the behavior of the oscillating neurons as they work to synchronize and the tension inherent to such a system, it is not set up to answer evolutionary questions. The fixed descriptions of the payoff parameters~\cite{EKD,yang2018kuramoto}, which are restricted to the type of Prisoner's Dilemma games, do not allow for the study of the explicit payoff conditions under which synchronization is most likely to occur. As the payoff is determined by numerical solutions to Kuramoto systems, it is intractable to derive closed-form conditions for natural selection to favor synchronization across a wide variety of scenarios.

To address this issue, we allow the payoff parameters a range of possible values, subject to a few biologically plausible assumptions (see details in Sec.~\ref{model}). This allows us to work backwards and discover what values of the payoff parameters cause the system to evolve into a state of synchronization, which in turn allows us to make inferences about the biology of the mammalian SCN. Additionally, while the framework of \cite{EKD} allows for the synchronization process to occur separately from the evolutionary dynamics of the neurons' strategies, the computational cost of accurately solving Kuramoto systems over long time frames is high for large populations of oscillators. In light of this, we consider intrinsic phases of neurons and their communicative strategies as combined traits that jointly determine their payoffs and are subject to natural selection. This novel setup leads to co-evolutionary dynamics of communicative strategies and multiple discrete phases of neurons, thereby providing a framework for a new model of coupled oscillatory systems that allows us to study the impact of the evolutionary constraint on the population of oscillators.  Consequently, we will be able to determine what evolutionary constraints allow organisms develop this synchronized oscillatory behavior. 

In this paper, we study the simplest case where oscillators either communicate with all other oscillators or none. We define a game between neurons, also inspired by the Prisoner's Dilemma~\cite{doebeli2005models}, in which each neuron receives a payoff based on their current level of synchrony with their neighboring neurons and whether or not they choose to communicate to improve the synchrony of the region. Using standard techniques from evolutionary games in finite populations~\cite{nowak2004emergence,Antal2006fixation,FUDENBERG2006imitation,traulsen2008analytical}, we are able to determine when communication is a favorable strategy for the population. We find that, under a variety of assumptions, this choice to communicate -- and thus synchronize -- is favored when the benefit received by two synchronized, communicating neurons exceeds twice the neuron's incurred cost of communication. 

\section{Model}
\label{model}
We consider a population of $n$ neurons, thought of as oscillating agents. We construct the biologically-motivated game under the following assumptions:
\begin{itemize}
\item neurons benefit by being in synchronization with their neighbors: the closer to synchronization, the greater the benefit; 
\item to influence one another, neurons must communicate with their neighbors, which incurs a cost.
\end{itemize}
Each neuron's strategy consists of the pair of their communicative state, a `$C$' if they communicate with their neighbors or an `$N$' if not, and their phase $\phi_j = j\frac{2\pi}{d}$, where $\phi_1$ through $\phi_d$ are distributed on the circle. The payoff of a neuron in any particular round of the game depends on its communicative state, the communicative state of its partner, and the cyclic difference in their phases, denoted $\Delta\phi$\footnote{If one neuron has phase $\phi_j$ and another neuron has phase $\phi_k$, $\Delta\phi = |\phi_j-\phi_k|$ if  $|\phi_j-\phi_k| \le \pi$, otherwise $\Delta\phi = 2\pi - |\phi_j-\phi_k| $.}. Thus, the payoff matrix, which describes pairwise interactions between neurons, takes on one of three possible forms, depending on their communicative strategies and phases:


\begin{eqnarray}
\begin{array}{c|cc}
(I) & (C,\phi_j) & (C,\phi_k) \\\hline
(C,\phi_j) & B(0)-c & B(\Delta\phi)-c\\
(C,\phi_k) & B(\Delta\phi)-c & B(0)-c
\end{array},\\
\begin{array}{c|cc}
(II) & (C, \phi_j) & (N, \phi_k) \\\hline
(C, \phi_j) & B(0)-c & \beta(\Delta\phi)-c\\
(N, \phi_k) & \beta(\Delta\phi) & 0
\end{array},\\
\begin{array}{c|cc}
(III) & (N,\phi_j) & (N,\phi_k) \\\hline
(N,\phi_j) & 0 & 0\\
(N,\phi_k) & 0 & 0 
\end{array}
\end{eqnarray}


\noindent Here, $c$ represents the cost of communication, while $B(\Delta\phi)$ is the benefit received by a communicative neuron playing against another communicative neuron and $\beta(\Delta\phi)$ is the benefit received by each neuron when only one is communicative. To incorporate the above assumptions, we take both $B$ and $\beta$ to be decreasing functions of $\Delta\phi$ and assume that the benefit of bilateral communication $B(\Delta\phi)$ is greater than that of unilateral communication $\beta(\Delta\phi)$, namely, $B(\Delta\phi) > \beta(\Delta\phi)$ for all $\Delta\phi$. 

While the payoff matrices above allow for pairwise comparisons of the relative strengths of various strategies, other quantities allow for more overarching comparisons. The \textit{expected payoff} of a strategy $E$, denoted by $\pi_E$, calculates the average payoff received by strategy $E$ given the current frequency of each strategy among its neighbors. The \textit{fitness} of a neuron with strategy $E$ is given by $f_E = e^{\delta\pi_E}$, where the parameter $\delta$ is the strength of selection \cite{traulsen2008analytical}. The fitness of a neuron's strategy is used to weigh the neuron's ability to reproduce in the evolution of the population, thereby mimicking the effect of natural selection on advantageous traits in biological evolution. 

The evolution of the system is governed by the Moran process: at each time step, we choose a neuron uniformly at random to delete, and one neuron is chosen with probability proportional to its selective fitness to reproduce, thereby replacing the deleted neuron with a new neuron of its own strategy \cite{evol_dynamics}. To better mimic biological evolution, we allow the strategy of a new neuron the chance to mutate during each reproduction step, meaning that with a small probability $\mu$, the new neuron will be assigned a random strategy rather than faithfully inherit the strategy of its parent. The strength of selection, $\delta$, determines the extent to which the structure of the game impacts the evolutionary success of each strategy: large values of $\delta$ give more weight to the role of payoff in fitness, while a selection strength of $\delta = 0$ gives all individuals the baseline fitness value of 1. The latter process is called \textit{neutral drift} \cite{evol_dynamics}, since the game plays no role in an individual's reproductive success. Thus, any neuron with a strategy with a higher average payoff and thus a higher fitness will be more likely to reproduce at each time step in the evolutionary process. Over time, we expect strategies with higher fitness values to increase in number in the population. Eventually, the Moran process will reach equilibrium in a state where all neurons share the same strategy \cite{evol_dynamics}. Once reaching such an equilibrium point, the state of the system will remain unchanged, unless a mutation event occurs.

Synchronization will usually\footnote{The one highly unlikely exception is the case where all neurons have the strategy $(N,\phi_j)$ for some phase $\phi_j$.} only occur when all neurons have a $(C, \star)$ strategy: regardless of the assumed underlying network topology, any neuron with an $(N,\star)$ strategy is effectively unconnected, as it does not take advantage of its links to neighboring nodes for communication -- and thus synchronization -- purposes. Thus, in our effort to determine conditions under which the system will synchronize, we look for those conditions under which $(C,\star)$ strategies are selectively favored. In this work, we assume the simplest possible underlying network topology (the so-called well-mixed populations~\cite{nowak2004emergence}): all neurons are equally likely to interact with all other neurons in the population. 

There are two standard simplifying assumptions in the evolutionary game theory literature: weak selection (when $\delta \ll \frac{1}{n}$)~\cite{nowak2004emergence,Antal2009mutation} and low mutation (when $\mu \ll \frac{1}{n}$)~\cite{FUDENBERG2006imitation,wu2012small}. In the results below, we explore the conditions under which communicative strategies are favored under various combinations of these two standard assumptions. Mathematically, problems are usually most theoretically tractable when both weak selection and low mutation assumptions are made, and thus we began with this case in Section \ref{Sec:WeakSel_LowMut} for pairwise invasion dynamics in the limit of weak selection, followed by the case where we assume instead that selection is strong in Section \ref{Sec:StrSel_LowMut}. While more difficult for studying evolutionary dynamics of multiple types, weakening only one of these assumptions can also often lead to tractable problems, as explored in Section \eqref{Sec:AnySel_LowMut} for the low mutation limit and Section \eqref{Sec:WeakSel_AnyMut} for the weak selection limit. Abandoning both often makes it quite difficult to obtain theoretical results, and thus we reserve such explorations for future work.

\section{Results}
\label{res}
To derive analytical approximation results, we consider this framework under various scenarios: (1) pairwise invasion dynamics where the population essentially consists of at most two different types of neurons at a time for both weak selection and strong selection limits, (2) evolutionary dynamics of multiple $2d$ types of neurons for any selection and low mutation, as well as for weak selection and any mutation. In each scenario, we are able to apply existing evolutionary game theoretic techniques to explore conditions under which communicative strategies are favored. 

\subsection{Pairwise invasion dynamics: weak selection limit} 
\label{Sec:WeakSel_LowMut}
Under the assumptions of weak selection $\delta \ll 1/n$, we show that communicative strategies benefit from selective pressure, meaning the fixation probability of a single communicative neuron $\rho_C$ exceeds the neutral fixation probability $1/n$, when $B(0) + \beta(\Delta\phi) > 3c$, and that they are favored over non--communicative strategies, that is, $\rho_C > \rho_N$, when $B(0) > 2c$. 

\begin{figure}
     \centering
     \includegraphics[width=\columnwidth]{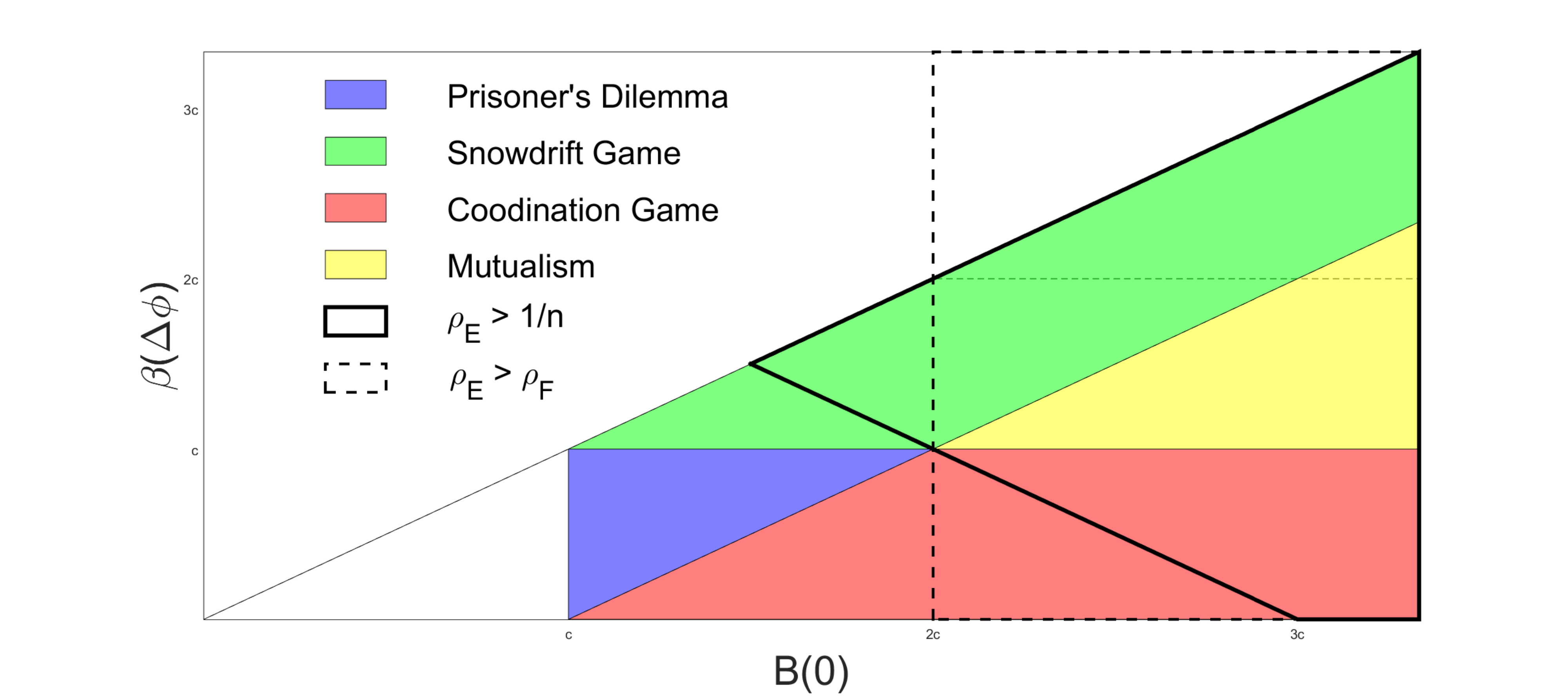}
     \caption{Classical game regimes for different values of $B(0)$ and $\beta(\Delta \phi)$ under the assumption of weak selection and low mutation. Also marked are regions where $E = (C,\phi_j)$ has a selective advantage, either over a neutral strategy (heavy line) or over the strategy $F = (N,\phi_k)$ (dotted line).}
     \label{fig:game_type}
\end{figure}

Assuming pairwise invasion dynamics means that if we start from a uniform--strategy state, only one mutation event occurs before the chain is re--absorbed into either its original uniform--strategy state or a state where the newly--mutated strategy is now universal. Thus, at most two strategies, $E$ and $F$, exist in the population at any given time. In this section, we will focus on the case where the two strategies present are $(C,\phi_j)$ and $(N,\phi_k)$ and ignore the cases where the two strategies existing in the population are $(C, \phi_j)$ and $(C, \phi_k)$ or $(N, \phi_j)$ and $(N, \phi_k)$. In these latter two cases, not only are the dynamics well understood (when both strategies present in the population are communicative, payoff matrix (I) above shows that the dynamics are driven by a simple coordination game, while payoff matrix (III) shows that in the non--communicative case, dynamics will be drive by the neutral process), but without a mix of $(C,\star)$ and $(N,\star)$ strategies, it is not possible to compare the relative success of communication and non--communication, which is the main goal of this work. Thus, in this section, we will restrict our study to the case where $E = (C,\phi_j)$  and $F = (N,\phi_k)$. In this setting, we can characterize the game being played in terms of classical games -- prisoner's dilemma, snowdrift game, cooperation, mutualism -- via the payoff matrix (II) above \cite{evol_dynamics}. Figure \eqref{fig:game_type} shows which of these classical games arise in various parameter regimes. 

In any population with only two strategies, we model the Moran process as a Markov chain with transition matrix $p$ on state space $\{0, 1, \dots, n-1, n\}$, where the chain is in state $i$ if there are $i$ individuals with strategy $E$ in the population. The transition matrix $p$ is a tri--diagonal stochastic matrix where\footnote{The low mutation assumption allows the transition probabilities below to ignore the possibility of a transition arising by mutation. Since, beginning from a uniform--strategy state, only one mutation event occurs before the chain is reabsorbed, there is no possibility that a mutation will arise during this transient phase of the process.}

\begin{align}
    &p_{0, 0} = 1, \\
    &p_{n,n} = 1, \\
    &p_{i,i-1} = \frac{i}{n}\cdot\frac{(n-i)f_F(i)}{if_E(i) + (n-i)f_F(i)}, \label{WSLMeqn:p_i,i-1}\\
    &p_{i,i+1} = \frac{n-i}{n}\cdot\frac{if_E(i)}{if_E(i) + (n-i)f_F(i)}, \text{ and} \label{WSLMeqn:p_i,i+1}\\
    &p_{i,i} = 1 - p_{i,i-1} - p_{i,i+1}.  \label{WSLMeqn:p_i,i}
\end{align}

From this, we derive the fixation probability, which allows us to determine the relative evolutionary success of various strategies. The \textit{fixation probability} of strategy $E = (C,\phi_j)$, denoted $\rho_E$, is the probability that, in a population with one strategy--$E$ individual and $n-1$ individuals with strategy $F = (N,\phi_k)$, the process is absorbed into the state where all individuals have strategy $E$ \cite{evol_dynamics}. Thus, we can quantify the success of a strategy $E$ in two ways: 

\begin{enumerate}
    \item Is $E$ more likely to fixate in the population than a strategy would under the neutral process, and 
    \item Is $E$ more likely to fixate in the population than $F$, its competing strategy? \cite{nowak2004emergence}  
\end{enumerate}

\noindent To answer both questions, the key quantity turns out to be the ratio $\frac{p_{i,i-1}}{p_{i,i+1}}$, which we denote $\gamma_i$. This is perhaps not surprising: if $\gamma_i$ is greater than 1, communicative strategies should be favored, while non--communicative strategies should be favored otherwise. We will see this quantity appear repeatedly in our various analyses under each set of assumptions.

We first compute the expected payoffs for both strategy $E$ and strategy $F$ as a function of the total number of strategy--$E$ individuals in the population, $i$, by summing the product of each potential payoff by the frequency with which such a neuron would receive that payoff. We get

\begin{align}
    \pi_E(i) &= \Big(B(0)-c\Big)\Bigg(\frac{i-1}{n-1}\Bigg) + \Big(\beta(\Delta\phi)-c\Big)\Bigg(\frac{n-i}{n-1}\Bigg) \nonumber \\
    &= \frac{1}{n-1}\bigg(i\Big(B(0) - \beta(\Delta\phi)\Big) + n\beta(\Delta\phi) - B(0) - (n-1)c\bigg),
\end{align}

\noindent and

\begin{align}
    \pi_F(i) &= \Big(\beta(\Delta\phi)\Big)\Bigg(\frac{i}{n-1}\Bigg) + \Big(0\Big)\Bigg(\frac{n-i-1}{n-1}\Bigg) \nonumber \\
    &= \beta(\Delta\phi)\Bigg(\frac{i}{n-1}\Bigg).
\end{align}

Next, we compute the probability, $x_i$, that the chain is absorbed into state $n$ from the state $i$. With this notation, we have $\rho_E = x_1$ and $\rho_F = 1- x_{n-1}$. We first observe the following recurrence:

\begin{align} 
    &x_0 = 0 \nonumber \\
    &x_i = p_{i,i-1}x_{i-1} + p_{i,i}x_i + p_{i,i+1}x_{i+1} \label{WSLMeqn:x_i}\\
    &x_n = 1. \nonumber 
\end{align}

\noindent If we let $y_i = x_i - x_{i-1}$ for $i = 1, \dots, n$, we have that

\begin{align}
    \sum_{i=1}^n y_i &= (x_1 - x_0) + (x_2 - x_1) + \dots + (x_n - x_{n-1}) \nonumber\\
    &= x_n - x_0 \nonumber \\
    &= 1. \label{eqn:sum_yi} 
\end{align}

\noindent Taking equation \eqref{WSLMeqn:x_i} with equation \eqref{WSLMeqn:p_i,i}, we get

\begin{align}
    p_{i,i-1}(x_i - x_{i-1}) &= p_{i,i+1}(x_{i+1} - x_i) \nonumber \\
    \frac{p_{i,i-1}}{p_{i,i+1}}y_i &= y_{i+1} \nonumber \\
    \gamma_i y_i &= y_{i+1}.  
\end{align}

\noindent Here, we see the quantity $\gamma_i$ appear for the first time. Now, since $y_1 = x_1$, we have $y_i = \prod_{j=1}^{i-1}\gamma_jx_1$ for $i \geq 2$. Substituting this into equation \eqref{eqn:sum_yi}, we get

\begin{align}
    &x_1 + \sum_{i=2}^n \prod_{j=1}^{i-1} \gamma_j x_1 = 1 \nonumber \\
    &x_1 \Bigg(1 + \sum_{i=1}^{n-1} \Big(\prod_{j=1}^i \gamma_j \Big) \Bigg) = 1 \nonumber \\
    &x_1 = \frac{1}{1 + \sum_{i=1}^{n-1} \Big(\prod_{j=1}^i \gamma_j \Big)}. 
\end{align}

\noindent Moreover, $x_i = \sum_{j=1}^i y_j = x_1\big(1 + \sum_{j=1}^{i-1}\prod_{k=1}^j\gamma_k \big)$. So:

\begin{equation} \label{eqn:x_i_full}
    x_i = \frac{1 + \sum_{j=1}^{i-1}\prod_{k=1}^j\gamma_k}{1 + \sum_{j=1}^{n-1} \Big(\prod_{k=1}^j \gamma_k \Big)}.
\end{equation}

\noindent Thus, we get that 

\begin{equation} \label{WSLMeqn:rho_E_gen}
    \rho_{E} = x_1 = \frac{1}{1 + \sum_{j=1}^{n-1} \Big(\prod_{k=1}^j \gamma_k \Big)},
\end{equation}

\noindent and

\begin{align}
    \rho_{F} &= 1 - x_{n-1} \nonumber \\
    &= 1 - \frac{1 + \sum_{j=1}^{n-2}\prod_{k=1}^j\gamma_k}{1 + \sum_{j=1}^{n-1} \Big(\prod_{k=1}^j \gamma_k \Big)} \nonumber \\
    &= \frac{\prod_{k=1}^{n-1} \gamma_k}{1 + \sum_{j=1}^{n-1} \Big(\prod_{k=1}^j \gamma_k \Big)} \nonumber \\
    &= \rho_E\prod_{k=1}^{n-1}\gamma_k\label{WSLMeqn:rho_F_gen}. 
\end{align}

To apply the general formulas \eqref{WSLMeqn:rho_E_gen} and \eqref{WSLMeqn:rho_F_gen} to our set-up, we first compute $\gamma_k$ with the appropriate values from payoff matrix (II):

\begin{align}
    \gamma_k &= \frac{p_{k,k-1}}{p_{k,k+1}} \nonumber \\
    &= \frac{\frac{k}{n}\cdot\frac{(n-k)f_F(k)}{kf_E(k) + (n-k)f_F(k)}}{\frac{n-k}{n}\cdot\frac{kf_E(k)}{kf_E(k) + (n-k)f_F(k)}} \nonumber \\
    &= \frac{f_F(k)}{f_E(k)} \nonumber \\
    &= \frac{e^{\delta\pi_F(k)}}{e^{\delta\pi_E(k)}} \nonumber \\ 
    &= e^{\delta\big(\pi_F(k) - \pi_E(k)\big)} \nonumber \\
    &= e^{\delta\Big(\beta(\Delta\phi)\big(\frac{k}{n-1}\big) - \frac{1}{n-1}\Big(k\big(B(0) - \beta(\Delta\phi)\big) + n\beta(\Delta\phi) - B(0) - (n-1)c\Big)\Big)} \nonumber \\
    &= e^{\frac{\delta}{n-1}\Big(\big(2\beta(\Delta\phi) - B(0)\big)k + B(0) - n\beta(\Delta\phi) + (n-1)c\Big)}. \label{WSLMeqn:gamma}
\end{align}

\noindent Thus, equations \eqref{WSLMeqn:rho_E_gen} -- \eqref{WSLMeqn:gamma} give us

\begin{align}
    \rho_{E} &= \Bigg(1 + \sum_{j=1}^{n-1}\prod_{k=1}^j e^{\frac{\delta}{n-1}\Big(\big(2\beta(\Delta\phi) - B(0)\big)k + B(0) - n\beta(\Delta\phi) + (n-1)c\Big)}\Bigg)^{-1} \nonumber \\
    &= \Bigg(1 + \sum_{j = 1}^{n-1} e^{\frac{\delta}{n-1}\sum_{k = 1}^j \big(2\beta(\Delta\phi) - B(0)\big)k + B(0) - n\beta(\Delta\phi) + (n-1)c\Big)}\Bigg)^{-1}  \nonumber \\
    &= \Bigg(1 + \sum_{j = 1}^{n-1} e^{\frac{\delta}{n-1}\Big(j^2\big(\beta(\Delta\phi) - \frac{1}{2}B(0)\big) + j\big(\frac{1}{2}B(0) - (n-1)\beta(\Delta\phi) + (n-1)c\big)\Big)}\Bigg)^{-1}, \label{WSLMeqn:rho_E}
\end{align}

\noindent and

\begin{align}
    \rho_F &= \rho_E\prod_{k=1}^{n-1} e^{\frac{\delta}{n-1}\Big(\big(2\beta(\Delta\phi) - B(0)\big)k + B(0) - n\beta(\Delta\phi) + (n-1)c\Big)} \nonumber \\
    &= \rho_E\bigg(e^{\frac{\delta}{n-1}\sum_{k=1}^{n-1}\big(\big(2\beta(\Delta\phi) - B(0)\big)k + B(0) - n\beta(\Delta\phi) + (n-1)c\big)}\bigg) \nonumber \\
    &= \rho_E\bigg(e^{\frac{\delta}{n-1}\big(\big(2\beta(\Delta\phi) - B(0)\big)\frac{(n-1)n}{2} + (n-1)\big(B(0) - n\beta(\Delta\phi) + (n-1)c\big)\big)}\bigg) \nonumber \\
    &= \rho_E\bigg(e^{\delta\big((n-1)c - \frac{n-2}{2}B(0)\big)}\bigg). \label{WSLMeqn:rho_F}
\end{align}

We are now prepared to answer the two questions posed above. For question 1, we first calculate the fixation probability for a strategy under the neutral process, i.e. when $\delta = 0$. Using equation \eqref{WSLMeqn:gamma} with $\delta = 0$ yields $\gamma_k = 1$ regardless of the particular game theoretic framework. From equation \eqref{WSLMeqn:rho_E_gen}, we then see that the fixation probability of any strategy under the neutral process is $\frac{1}{n}$. Thus, to determine whether the selective pressure on strategy $E = (C,\phi_j)$ increases the overall evolutionary success of the strategy, we compare equation $\eqref{WSLMeqn:rho_E}$ to $\frac{1}{n}$. Here, we take advantage of our assumption of weak selection ($\delta << \frac{1}{n}$) and use a first order Taylor expansion in $\delta$ to approximate $\gamma_k$ (see equation \eqref{WSLMeqn:gamma}):

\begin{align}
    \gamma_k &\approx 1 + \delta\big(\pi_F(k) - \pi_E(k)\big) \nonumber \\
    &= 1 + \frac{\delta}{n-1}\Big(\big(2\beta(\Delta\phi)-B(0)\big)k + B(0) - n\beta(\Delta\phi) + (n-1)c\Big) \label{WSLMeqn:gamma_approx}
\end{align}

\noindent Substituting this expression into equation \eqref{WSLMeqn:rho_E_gen} allows us to approximate $\rho_E$ as follows:
\small
\bgroup
\begin{align}
    \rho_E &\approx \Bigg(1 + \sum_{j=1}^{n-1} \Bigg(\prod_{k=1}^j \bigg(1 + \frac{\delta}{n-1}\Big(\big(2\beta(\Delta\phi)-B(0)\big)k + B(0) - n\beta(\Delta\phi) + (n-1)c\Big)\bigg) \Bigg)\Bigg)^{-1} \nonumber \\
    &\approx \Bigg(1 + \sum_{j=1}^{n-1}\bigg(1 + \frac{\delta}{n-1}\sum_{k=1}^j \Big(\big(2\beta(\Delta\phi) - B(0)\big)k + B(0) - n\beta(\Delta\phi) + (n-1)c\Big)\bigg)\Bigg)^{-1} \nonumber \\
    &= \Bigg(1 + \sum_{j=1}^{n-1}\bigg(1 + \frac{\delta}{n-1}\Big(\big(2\beta(\Delta\phi) - B(0)\big)\frac{j(j+1)}{2} + j\big(B(0) - n\beta(\Delta\phi) + (n-1)c\big)\Big)\bigg)\Bigg)^{-1} \nonumber \\
    &= \Bigg(1 + \sum_{j=1}^{n-1}\bigg(1 + \frac{\delta}{n-1}\Big(j^2\Big(\beta(\Delta\phi) - \frac{B(0)}{2}\Big) + j\Big(\frac{B(0)}{2} - (n-1) \beta(\Delta\phi) + (n-1)c)\Big)\Big)\bigg)\Bigg)^{-1} \nonumber \\
   &= \Bigg(n + \frac{\delta}{n-1}\bigg(\Big(\beta(\Delta\phi) - \frac{1}{2}B(0)\Big)\frac{(n-1)n(2n-1)}{6} \nonumber \\
   &\hspace{2cm} + \Big(\frac{1}{2}B(0) - (n-1)\beta(\Delta\phi) + (n-1)c\Big)\frac{(n-1)n}{2}\bigg)\Bigg)^{-1} \nonumber \\
   &= \Bigg(n\bigg(1 + \delta\bigg(\big(\beta(\Delta\phi) - \frac{1}{2}B(0)\big)\frac{2n-1}{6} + \Big(\frac{1}{2}B(0) - (n-1)\beta(\Delta\phi) + (n-1)c\Big)\frac{1}{2}\bigg)\bigg)\Bigg)^{-1} \nonumber \\
   &= \frac{1}{n}\Bigg(1 - \frac{\delta}{6}\bigg(\Big(B(0) + \beta(\Delta\phi) - 3c\Big)n - \Big(2B(0) + 2\beta(\Delta\phi) - 3c\Big)\bigg)\Bigg)^{-1} 
\end{align}
\egroup
\normalsize

\noindent Letting $\alpha = B(0) + \beta(\Delta\phi) - 3c$ and $\lambda = 2B(0) + 2\beta(\Delta\phi) - 3c$, we have 

\begin{equation}
    \rho_E \approx \frac{1}{n}\cdot\frac{1}{1 - \frac{\delta}{6}\big(\alpha n - \lambda\big)}.
\end{equation}

\noindent Thus, $\rho_E$ will exceed $\frac{1}{n}$ if and only if $\alpha n > \lambda$, or rather

\begin{align}
    n\big(B(0) + \beta(\Delta\phi) - 3c\big) &> 2B(0) + 2\beta(\Delta\phi) - 3c \nonumber \\
    (n-2)B(0) + (n-2)\beta(\Delta\phi) &> 3c(n-1).  \label{WSLMcond:neut_fin_pop}
\end{align}

\noindent For sufficiently large populations, we then get the following simpler condition that specifies exactly when strategy $E$ has greater evolutionary success than a strategy under the neutral process:

\begin{equation} \label{WSLMcond:neut}
    B(0) + \beta(\Delta\phi) > 3c.
\end{equation}

We now turn our attention to question (2): is strategy $E = (C,\phi_j)$ more likely to fixate in the population than strategy $F = (N,\phi_k)$? To examine this, we look at the ratio of $\rho_F$ to $\rho_E$, using equation \eqref{WSLMeqn:rho_F}:

\begin{align}
    \frac{\rho_F}{\rho_E} &= \frac{\rho_E\bigg(e^{\delta\big((n-1)c - \frac{n-2}{2}B(0)\big)}\bigg)}{\rho_E} \nonumber \\
    &= e^{\delta\big((n-1)c - \frac{n-2}{2}B(0)\big)}. \label{WSLMeqn:fix_prob_ratio}
\end{align}

\noindent Thus, $\rho_E > \rho_F$ if and only if 

\begin{equation} \label{WSLMcond:rhoE_2_rhoF_fin_pop}
    (n-1)c < \frac{n-2}{2}B(0).
\end{equation}

\noindent Again, for sufficiently large populations, we get the simpler condition

\begin{equation} \label{WSLMcond:rhoE_2_rhoF}
    B(0) > 2c.
\end{equation}

\noindent In Section \ref{Sec:AnySel_LowMut} below, we will prove that this condition holds even under slightly weaker conditions.

\subsection{Pairwise invasion dynamics: strong selection limit} 
\label{Sec:StrSel_LowMut}
Under the assumptions of strong selection, we show that communicative strategies are always favored over non--communicative strategies in pairwise invasion dynamics when $c < \beta(\Delta\phi) < B(0) - c$. 

If we instead assume that selection is strong ($\delta >> 1$), there is no longer a useful way to estimate a strategy's fixation probability, so we take a different approach. We instead consider the critical ratio $\gamma_i^{-1} = \frac{p_{i,i+1}}{p_{i,i-1}}$ directly, for $1 \leq i \leq n-1$. As mentioned in Section \ref{Sec:WeakSel_LowMut} above, for a given $i$, this ratio provides an indication of whether the number of strategy--$E$ individuals is more likely to increase or decrease: if $\gamma_i^{-1} < 1$, the population is more like to lose a strategy--$E$ individual, while if $\gamma_i^{-1} > 1$, the population is more likely to gain a strategy--$E$ individual. Recall from equation \eqref{WSLMeqn:gamma} that 

\begin{align}
    \gamma_i^{-1} &= \frac{p_{i,i+1}}{p_{i,i-1}} \\\nonumber
    &= \Big(e^{\delta(\pi_F(i) - \pi_E(i))}\Big)^{-1} \\\nonumber
    &= e^{\delta(\pi_E(i) - \pi_F(i))}.
\end{align}

\noindent Thus, in the limit of strong selection (i.e. as $\delta \rightarrow \infty$), $\gamma_i^{-1}$ approaches either $\infty$, if $\pi_E(i) - \pi_F(i) > 0$, or $0$, if $\pi_E(i) - \pi_F(i) < 0$. Therefore, under the assumption of strong selection the process becomes essentially deterministic: whether or not the process moves to a state with a higher or lower number of strategy--$E$ individuals is determined by the sign of $\pi_E(i) -  \pi_F(i)$. Furthermore, $\pi_E(i) - \pi_F(i)$, which we'll denote as $\Delta\pi(i)$, is a linear function of $i$:

\begin{align}
    \Delta\pi(i) &= \pi_E(i) - \pi_F(i) \nonumber \\
    &= \Bigg(\frac{1}{n-1}\Big(i\big(B(0) - \beta(\Delta\phi)\big) + n\beta(\Delta\phi) - B(0) - (n-1)c\Big)\Bigg) - \Bigg(\beta(\Delta\phi)\bigg(\frac{i}{n-1}\bigg)\Bigg) \nonumber \\
    &= \frac{1}{n-1}\Big((B(0) - 2\beta(\Delta\phi))i + (n\beta(\Delta\phi) - B(0) - (n-1)c)\Big). \label{SSLMeqn:pi_diff}
\end{align}

\noindent We can characterize the dynamics of the system simply by considering this difference at the end points, where $i = 1$ and $i = n-1$:

\begin{align}
    \Delta\pi(1) &= \frac{1}{n-1}\Big(\beta(\Delta\phi)(n-2) - c(n-1)\Big), \nonumber \\
    \Delta\pi(n-1) &= \frac{1}{n-1}\Big(B(0)(n-2)-\beta(\Delta\phi)(n-2) - c(n-1)\Big).
\end{align}

\noindent Thus, when there are few communicative strategy individuals in the population, communication is favored when

\begin{align}
    \Delta\pi(1) &> 0 \nonumber \\
    \frac{1}{n-1}\Big(\beta(\Delta\phi)(n-2) - c(n-1)\Big) &> 0 \nonumber \\
    \beta(\Delta\phi)(n-2) &> c(n-1) \nonumber \\
    \beta(\Delta\phi) &> \frac{c(n-1)}{n-2}. \label{SSLMcond:C_favor}
\end{align}

\begin{figure}
    \centering
    \includegraphics[width=6in,height=4.5in]{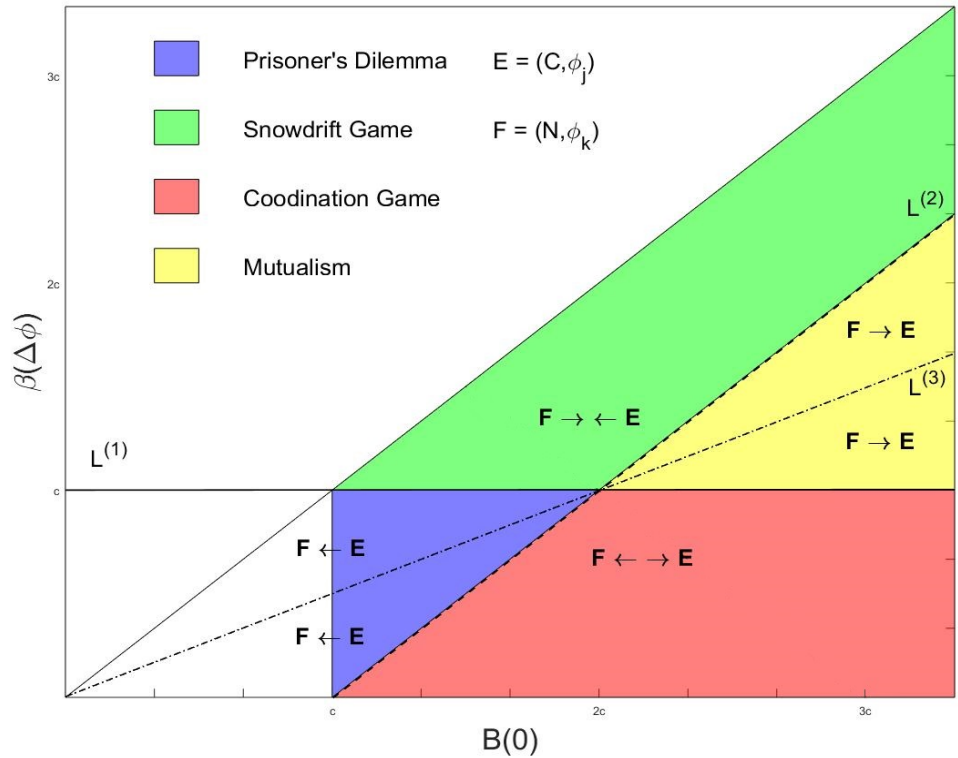}
    \caption{Game regimes for different values of $B(0)$ and $\beta(\Delta\phi)$ in the case of strong selection. Above lines $L^{(1)}$ and $L^{(2)}$, strategy $F$ is favored when there are more strategy--$E$ individuals, and strategy $E$ is favored when there are more strategy--$F$ individuals. Below $L^{(2)}$ and above $L^{(1)}$, strategy $E$ is always favored, while strategy $F$ is always favored in the region below $L^{(1)}$ and above $L^{(2)}$. Finally, below both lines, strategy $E$ is favored when there are more strategy--$E$ individuals, and strategy $F$ is favored when there are more strategy--$F$ individuals in the population.}
    \label{SSLMfig:dynamic_reg}
\end{figure}

\noindent In a sufficiently large population, equation \eqref{SSLMcond:C_favor} implies that communication is favored when 

\begin{equation}\label{SSLMeqn:cond1}
    \beta(\Delta\phi) > c,
\end{equation}

\noindent which is depicted in Figure \eqref{SSLMfig:dynamic_reg} as the region above line $L^{(1)}$.

When there are few non--communicative strategy individuals in the population, communication is favored when

\begin{align}
    \Delta\pi(n-1) &> 0 \nonumber \\
    \frac{1}{n-1}\Big(B(0)(n-2) - \beta(\Delta\phi)(n-2) - c(n-1)\Big) &> 0 \nonumber \\
    \beta(\Delta\phi)(n-2) &< B(0)(n-2) - c(n-1) \nonumber \\
    \beta(\Delta\phi) &< B(0) - \frac{c(n-1)}{n-2}.
\end{align}

\noindent Again, when the population is sufficiently large, this implies that communication is favored if and only if 

\begin{equation}\label{SSLMeqn:cond2}
    \beta(\Delta\phi) < B(0) - c,
\end{equation}

\noindent which is represented by the region below line $L^{(2)}$ in Figure \eqref{SSLMfig:dynamic_reg}. Together, conditions \eqref{SSLMeqn:cond1} and \eqref{SSLMeqn:cond2} divide the parameter space into regions where the dynamics differ, as seen in Figure \eqref{SSLMfig:dynamic_reg}.

\subsection{Evolutionary dynamics of multiple types: any selection strength and low mutation limit.} \label{Sec:AnySel_LowMut}
We prove that under the assumption of low mutation, $(C,\star)$ strategies are favored by the selection process if and only if $B(0) > 2c$. More specifically, we will show that the frequency of $(C,\star)$ strategies is higher than that of the $(N,\star)$ strategies at stationarity if and only if $B(0) > 2c$. 

Suppose we have $d$ phases, $\phi_1, \phi_2, \dots, \phi_d$ with $\phi_j = j\frac{2\pi}{d}$, evenly distributed on the cycle. We thus have $2d$ strategies: $(C, \phi_1)$, $(C, \phi_2)$, $\dots$, $(C, \phi_d)$, $(N, \phi_1)$, $(N, \phi_2)$, $\dots$, and $(N, \phi_d)$. As described above, if the mutation rate is sufficiently low, we assume that in any uniform--strategy population, only one mutation will occur before the chain is reabsorbed and thus that no more than two of these $2d$ total strategies ever exist in the population at any one time. Therefore, we can compress the Moran process described in Section \ref{Sec:WeakSel_LowMut} above into a new Markov chain that transitions between uniform--strategy states. This new chain has the set of possible strategies as its state space, and the probability of transitioning from, say, the uniform strategy--$F$ state to the uniform strategy--$E$ state is given by the probability that a single strategy--$E$ individual (arising by mutation) can overtake a population of all strategy--$F$ individuals \cite{Hauert2007via}. Thus, this transition probability is exactly the fixation probability, $\rho_{E}$, from equation \eqref{WSLMeqn:rho_E}. Here, we will consider this new Markov chain on the uniform--strategy states $\{(C, \phi_1), (C, \phi_2), \dots, (C, \phi_d), (N, \phi_1), (N, \phi_2), \dots, (N, \phi_d)\}$, where the transition probabilities are given by the corresponding fixation probabilities. The transition matrix, $M$, for this chain has a block structure

\begin{equation}
    M = \left[\begin{array}{ccc|ccc}
     & B_1 & & & B_2 & \\
     \hline
     & B_3 & & & B_4 & \\
     \end{array}\right],
\end{equation}

\noindent where $B_1$, $B_2$, $B_3$, and $B_4$ are $d$-by-$d$ blocks describing transitions between two $(C, \star)$ strategies, transitions from a $(C, \star)$ strategy to an $(N, \star)$ strategy, transitions from an $(N, \star)$ strategy to a $(C, \star)$ strategy, and transitions between $(N, \star)$ strategies, respectively. Diagonal elements in blocks $B_1$ and $B_4$ are assigned the values necessary to make $M$ row--stochastic\footnote{Strictly speaking, entries in the transition matrix $M$ are of the form $\mu\rho_{\_\_,\Delta\phi}$. Thus, by re--scaling the mutation probability $\mu$, we can ensure that diagonal entries of $M$ are, in fact, positive. Note that the factor $\mu$ has no bearing on the stationary distribution of $M$, and can thus be ignored \cite{FUDENBERG2006imitation}.}. Within each block, all fixation probabilities follow the same form. As we saw in equation \eqref{WSLMeqn:rho_E_gen}, these fixation probabilities depend only on the payoff values and the quantity $\Delta\phi$. Since the payoff values -- up to their own dependence on $\Delta\phi$ -- are constant within each block, the transition probabilities within a single block differ only by their dependence on $\Delta\phi$. Thus, to simplify notation and emphasize the important quantity $\Delta\phi$, we will write $\rho_{CN,\Delta\phi}$ to denote the fixation probability for a strategy $(N,\phi_j)$ invading a population of $(C, \phi_k)$ strategy individuals. 

These expressions are easily calculated using equation \eqref{WSLMeqn:rho_E_gen} and the payoff matrix associated to each block. For blocks $B_2$ and $B_3$, these fixation probabilities were previously calculated in Section \ref{Sec:WeakSel_LowMut} and can be found in equations \eqref{WSLMeqn:rho_F} and \eqref{WSLMeqn:rho_E}, respectively. For block $B_1$, substituting values from payoff matrix (I) into equation \eqref{WSLMeqn:rho_E_gen} yields the expression

\begin{equation}\label{ASLMeqn:CC_fix_prob}
    \rho_{CC,\Delta\phi} = \Bigg(1 + \sum_{k = 1}^{n-1} e^{\frac{\delta}{n-1}\big(\frac{k(k+1)}{2}(2B(\Delta\phi) - 2B(0)) + k(B(0)n - B(\Delta\phi)n)\big)}\Bigg)^{-1}.
\end{equation}

\noindent Finally, for block $B_4$, substituting values from payoff matrix (III) into equation \eqref{WSLMeqn:rho_E_gen} yields the constant value $\frac{1}{n}$. Given that all differences between transition probabilities in a single block are driven by $\Delta\phi$, these blocks exhibit many symmetries. For example, when $d$ is odd, we get that $B_1$ is given by

\begin{equation}
\resizebox{\columnwidth}{!}{$
    B_1 = \left[\begin{array}{ccccccccc}
    \dag & \rho_{CC,1} & \rho_{CC,2} & \dots & \rho_{CC,\lfloor\frac{d}{2}\rfloor} & \rho_{CC,\lfloor\frac{d}{2}\rfloor} & \rho_{CC,\lfloor\frac{d}{2}\rfloor-1} & \dots & \rho_{CC,1} \\
    \rho_{CC,1} & \dag & \rho_{CC,1} & \dots & \rho_{CC,\lfloor\frac{d}{2}\rfloor-1} & \rho_{CC,\lfloor\frac{d}{2}\rfloor} & \rho_{CC,\lfloor\frac{d}{2}\rfloor} & \dots & \rho_{CC,2} \\
    \rho_{CC,2} & \rho_{CC,1} & \dag & \dots & \rho_{CC,\lfloor\frac{d}{2}\rfloor-2} & \rho_{CC,\lfloor\frac{d}{2}\rfloor-1} & \rho_{CC,\lfloor\frac{d}{2}\rfloor} & \dots & \rho_{CC,3} \\
    \vdots & \vdots & \vdots & \ddots & \vdots & \vdots & \vdots & \ddots & \vdots \\
    \rho_{CC,\lfloor\frac{d}{2}\rfloor-1} & \rho_{CC,\lfloor\frac{d}{2}\rfloor-2} & \rho_{CC,\lfloor\frac{d}{2}\rfloor-3} & \dots & \dag & \rho_{CC,1} & \rho_{CC,2} & \dots & \rho_{CC,\lfloor\frac{d}{2}\rfloor} \\
    \rho_{CC,\lfloor\frac{d}{2}\rfloor} & \rho_{CC,\lfloor\frac{d}{2}\rfloor-1} & \rho_{CC,\lfloor\frac{d}{2}\rfloor-2} & \dots & \rho_{CC,1} & \dag & \rho_{CC,1} & \dots & \rho_{CC,\lfloor\frac{d}{2}\rfloor} \\
    \rho_{CC,\lfloor\frac{d}{2}\rfloor} & \rho_{CC,\lfloor\frac{d}{2}\rfloor} & \rho_{CC,\lfloor\frac{d}{2}\rfloor-1} & \dots & \rho_{CC,2} & \rho_{CC,1} & \dag & \dots & \rho_{CC,\lfloor\frac{d}{2}\rfloor-1} \\
    \vdots & \vdots & \vdots & \ddots & \vdots & \vdots & \vdots & \ddots & \vdots \\
    \rho_{CC,1} & \rho_{CC,2} & \rho_{CC,3} & \dots & \rho_{CC,\lfloor\frac{d}{2}\rfloor} & \rho_{CC,\lfloor\frac{d}{2}\rfloor-1} & \rho_{CC,\lfloor\frac{d}{2}\rfloor-2} & \dots & \dag \\
    \end{array}\right],%
$}
\end{equation}

\noindent where $\dag$ stands in for the values needed to ensure row--stochasticity of $M$. When $d$ is even, $B_1$ has the same general pattern, with the exception that, in each row and column, $\rho_{\_\hspace{.3mm}\_,\lfloor\frac{d}{2}\rfloor}$ is repeated only once.

This same pattern repeats itself in blocks $B_2$ and $B_3$, where $\rho_{CC,\Delta\phi}$ is replaced with $\rho_{CN,\Delta\phi}$ in $B_2$ and $\rho_{NC,\Delta\phi}$ in $B_3$. Furthermore, the diagonal values are replaced with $\rho_{CN,0}$ and $\rho_{NC,0}$ in blocks $B_2$ and $B_3$, respectively. Note that $\rho_{CN,\Delta\phi}$ and $\rho_{NC,\Delta\phi}$ are \textit{not} equal quantities, and $M$ is not a symmetric matrix.

However, the blocks $B_1$, $B_2$, $B_3$, and $B_4$ are each symmetric matrices. Within each of these blocks, each row (and thus each column) contains the same set of (non-diagonal) values, so all rows (and thus all columns) within a single block -- and thus for the matrix as a whole -- have the same sum, which implies that the diagonal elements of $B_1$ (the $\dag$ values) are all equal, as are the diagonal values in $B_4$. 

The symmetries present in $M$ allow us to show that the stationary distribution associated to this Markov chain has the form 

\begin{equation}
    \mathbf{s} = [\underbrace{s_1, s_1, \dots, s_1}_d, \underbrace{s_2, s_2, \dots, s_2}_d].
\end{equation}

\noindent In fact, we will prove that for $s_1 = \frac{\sum_{r=d+1}^{2d}M_{rq}}{d\sum_{r=d+1}^{2d}(M_{qr} + M_{rq})}$, $s_2 = \frac{\sum_{r=d+1}^{2d}M_{qr}}{d\sum_{r=d+1}^{2d}(M_{qr} + M_{rq})}$, and $1 \leq q \leq d$, $\mathbf{s}$ is the stationary distribution for $M$. While both $s_1$ and $s_2$ appear as if to vary with $q$, the symmetries of $M$ imply $s_1$ and $s_2$ are the same for $1 \leq q \leq d$. To see this, note that when $1\leq q \leq d$, $\sum_{r=d+1}^{2d} M_{rq}$ sums over a column of block $B_3$. Since all columns of $B_3$ contain the same set of values, choosing a different column does not change the sum. Similarly, when $1\leq q \leq d$, $\sum_{r=d+1}^{2d} M_{qr}$ is summing over a row in block $B_1$, in which all rows contain the same set of values. Thus, any choice of $q$ such that $1 \leq q \leq d$ will yield the same values for $s_1$ and $s_2$. Furthermore, $s_1$ and $s_2$ have the following relationship:

\begin{equation} \label{eqn:s1_and_s2}
    s_1\sum_{r=d+1}^{2d} M_{qr} = s_2\sum_{r=d+1}^{2d} M_{rq}.
\end{equation}

To prove that $\mathbf{s}$ is indeed the stationary distribution for $M$, we will show that $\mathbf{s}M = \mathbf{s}$. First, for $1 \leq u \leq d$, we have:

\begin{align}
    (\mathbf{s}M)_u &= \sum_{v=1}^{2d} \mathbf{s}_v M_{vu} \\
    &= \sum_{v=1}^d s_1 M_{vu} + \sum_{v=d+1}^{2d} s_2 M_{vu}.
\end{align}

\noindent Since $B_1$ is symmetric, we can switch the indices of $M$ in the first sum. Furthermore, we can use the relationship between $s_1$ and $s_2$ in equation \eqref{eqn:s1_and_s2} to substitute for the second sum:

\begin{align}
    (\mathbf{s}M)_u &= s_1\sum_{v=1}^d M_{uv} + s_1\sum_{v=d+1}^{2d} M_{uv} \\
    &= s_1 \sum_{v=1}^{2d} M_{uv} \\
    &= s_1,
\end{align}

\noindent where the final equality follows from the row stochasticity of $M$. Thus, we've shown that the first $d$ elements of $(\mathbf{s}M)$ are indeed $s_1$. A similar argument shows that the final $d$ elements are, in fact, $s_2$. If $d+1 \leq u \leq 2d$, we have:

\begin{align}
    (\mathbf{s}M)_u &= \sum_{v=1}^{2d} \mathbf{s}_vM_{vu} \nonumber \\
    &= \sum_{v=1}^d s_1 M_{vu} + \sum_{v=d+1}^{2d} s_2 M_{vu} \nonumber \\
    &= s_1\sum_{v=1}^d M_{u-d,v+d} + s_2\sum_{v=d+1}^{2d} M_{uv} \nonumber \\
    &= s_1\sum_{v=d+1}^{2d}M_{u-d,v} + s_2\sum_{v=d+1}^{2d}M_{uv},
\end{align}

\noindent where the second--to--last equality follows from the symmetry of blocks $B_2$ and $B_4$. Applying equation \eqref{eqn:s1_and_s2} and noting the symmetry of $B_3$, we get

\begin{align}
    (\mathbf{s}M)_u &= s_2\sum_{v=d+1}^{2d} M_{v,u-d} + s_2\sum_{v=d+1}^{2d}M_{uv} \nonumber \\
    &= s_2\sum_{v=d+1}^{2d} M_{u,v-d} + s_2\sum_{v=d+1}^{2d} M_{uv} \nonumber \\
    &= s_2\sum_{v=1}^d M_{uv} + s_2\sum_{v=d+1}^{2d} M_{uv} \nonumber \\
    &= s_2 \sum_{v=1}^{2d} M_{uv} \nonumber \\
    &= s_2.
\end{align}

\noindent Thus, $\mathbf{s}M = \mathbf{s}$. Since $\sum_{v=1}^{2d}\mathbf{s}_v = 1$, $\mathbf{s}$ is the stationary distribution for $M$.

With this stationary distribution in hand, we can compare frequencies of $(C,\star)$ and $(N,\star)$ strategies. First, though, note that the ratio given by equation \eqref{WSLMeqn:fix_prob_ratio},

\begin{equation}
    \frac{\rho_{CN,\Delta\phi_{qr}}}{\rho_{NC,\Delta\phi_{qr}}} = e^{\delta\big(\frac{n-2}{2}B(0) - (n-1)c\big)},
\end{equation} 

\noindent where $\Delta\phi_{qr}$ is the cyclic difference between phases $\phi_q$ and $\phi_r$, is constant for all possible pairs $(q,r)$. Thus, if we let $\omega = \linebreak e^{\delta\big(\frac{n-2}{2}B(0) - (n-1)c\big)}$, we have $\rho_{NC,\Delta\phi_{qr}} = \omega\rho_{CN,\Delta\phi_{qr}}$ for any $(q,r)$. 

At stationarity, $(C, \star)$ strategies occur with frequency $ds_1$, while $(N, \star)$ strategies occur with frequency $ds_2$. To compare, we look at the ratio:

\begin{align}
    \frac{d\cdot s_1}{d\cdot s_2} &= \frac{s_1}{s_2} \nonumber \\
    &= \frac{\sum\limits_{r=d+1}^{2d}M_{rq}}{\sum\limits_{r=d+1}^{2d}M_{qr}}.
\end{align}

\noindent Recalling that $1 \leq q \leq d$, we have

\begin{equation}
    \frac{d\cdot s_1}{d\cdot s_2} = \frac{\sum\limits_{r=1}^{d}\rho_{NC,\Delta\phi_{qr}}}{\sum\limits_{r=1}^{d}\rho_{CN,\Delta\phi_{qr}}}.
\end{equation}

\noindent Substituting for $\rho_{NC,\Delta\phi_{qr}}$ yields

\begin{align}
    \frac{d\cdot s_1}{d\cdot s_2} &= \frac{\sum\limits_{r=1}^d \omega\rho_{CN,\Delta\phi_{qr}}}{\sum\limits_{r=1}^d \rho_{CN,\Delta\phi_{qr}}} \nonumber \\
    &= \frac{\omega\sum\limits_{r=1}^d \rho_{CN,\Delta\phi_{qr}}}{\sum\limits_{r=1}^d \rho_{CN,\Delta\phi_{qr}}} \nonumber \\
    &= \omega \nonumber \\
    &= e^{\delta\big(\frac{n-2}{2}B(0) - (n-1)c\big)}.
\end{align}

\noindent Thus, for communicative strategies to be more prevalent at stationarity, we need 

\begin{equation}
    e^{\delta\big(\frac{n-2}{2}B(0) - (n-1)c\big)} > 1,
\end{equation}

\noindent which occurs exactly when 

\begin{equation}
    \frac{n-2}{2}B(0) - (n-1)c > 0,
\end{equation}

\noindent or rather when

\begin{equation}
    B(0) > \frac{2c(n-1)}{n-2}.
\end{equation}

\noindent For sufficiently large populations, it suffices to have $B(0) > 2c$. \qed

\subsection{Evolutionary dynamics of multiple types: weak selection limit and any mutation rate.} \label{Sec:WeakSel_AnyMut}
Under only the assumption of weak selection, we show that communicative strategies are favored over non--communicative strategies at any mutation rate when $$B(0) + \mu n\Bigg(\Big(\langle B(\Delta\phi_{l\cdot})\rangle - \frac{1}{2}\langle B(\Delta\phi_{\cdot\cdot})\rangle\Big) + \Big(\langle\beta(\Delta\phi_{l\cdot})\rangle - \langle\beta(\Delta\phi_{\cdot\cdot})\rangle\Big)-c\Bigg) > 2c,$$ where $\langle B(\Delta\phi_{l\cdot}) \rangle$, for example, denotes the average of $B(\Delta\phi_{lq})$ over all possible phases $\phi_q$. 

Without assuming low mutation, the population may consist of more than two strategies at any given time. However, following the framework in \cite{Antal2009mutation}, we study our multiple--strategy population under the assumption that the selective pressure is sufficiently weak (i.e. $\delta << \frac{1}{n}$). In that paper, Antal et al. identify a strategy's evolutionary success with its average frequency in the long--time average. Under the assumption of weak selection, all strategies have an approximately equal average frequency ($\frac{1}{m}$, where $m$ is the number of strategies) in the stationary distribution of the evolutionary process. Thus, a strategy is considered to be favored by selection if its long--time average frequency exceeds $\frac{1}{m}$. To make this more precise, suppose that our size $n$ population contains $m$ strategies, and let $A = (a_{qr})$ be the $m$--by--$m$ payoff matrix, where $a_{qr}$ is the payoff received by a strategy--$q$ player when playing a strategy--$r$ player. Once again, the evolutionary dynamics are given by the frequency--dependent Moran process, where a node's likelihood of reproduction is proportional to its frequency. The new node inherits its parent's strategy with probability $1-\mu$; with probability $\mu$, the node receives a random, newly--mutated strategy. Antal et al. state that strategy $l$ is favored by selection when 

\begin{equation}\label{WSAMcond:both_short}
    L_l + \mu nH_l > 0,
\end{equation}

\noindent where $L_l$ and $H_l$ (given below) are expressions that characterize the favorability of strategy $l$ when the mutation rate is low and high, respectively. The factor of $\mu n$ in front of $H_l$ allows the mutation rate to determine the relative importance of the low and high mutation terms. 

In \cite{Antal2009mutation}, then, if mutation is low ($\mu << \frac{1}{n}$), strategy $l$ is selectively advantaged when 

\begin{equation}\label{WSAMcond:Lk}
    L_l = \frac{1}{m}\sum_{q=1}^m (a_{ll} + a_{lq} - a_{ql} - a_{qq}) > 0.
\end{equation}

\noindent For intuition, note that, when the mutation rate is low and only two strategies, say $l$ and $q$, exist in the population, strategy $l$ is favored over strategy $q$ when $a_{ll} + a_{lq} - a_{ql} - a_{qq} > 0$ \cite{evol_dynamics}, which is consistent with condition \eqref{WSLMcond:rhoE_2_rhoF} as derived in Section \ref{Sec:WeakSel_LowMut}. Thus, here in this multi--strategy case, condition \eqref{WSAMcond:Lk} states that strategy $l$ has a selective advantage when the average of these critical values for each other strategy exceeds $0$.

Conversely, if the mutation rate is high, Antal et al. say strategy $l$ is favored when 

\begin{equation}\label{WSAMcond:Hk}
    H_l = \frac{1}{m^2}\sum_{q,r=1}^m (a_{lr} - a_{qr}) > 0.
\end{equation}

\noindent This condition is obtained by comparing the fitness of strategy $l$ to the average fitness of the population. Given our assumption that in the case of weak selection, all strategies have an approximately equal average frequency of $\frac{1}{m}$, the expected payoff for strategy $l$ is approximately $\frac{1}{m}\sum_{r=1}^m a_{lr}$. Thus, strategy $l$ has an approximate fitness of $1 + \frac{\delta}{m}\sum_{r=1}^m a_{lr}$.\footnote{Here, we approximate the fitness of strategy $l$ with its first--order Taylor approximation, as in equation \eqref{WSLMeqn:gamma_approx}.} Similarly, the average fitness for any strategy can then be expressed as $\frac{1}{m}\sum_{q=1}^m\big(1 + \frac{\delta}{m}\sum_{r=1}^m a_{qr}\big)$. Comparing these two expressions, we get the following:

\begin{align}
    \Bigg(1 + \frac{\delta}{m}\sum_{r=1}^m a_{lr}\Bigg) - \Bigg(\frac{1}{m}\sum_{q=1}^m\Big(1 + \frac{\delta}{m}\sum_{r=1}^m a_{qr}\Big)\Bigg) &= 1 + \frac{\delta}{m}\sum_{r=1}^m a_{lr} - 1 - \frac{\delta}{m^2}\sum_{q,r=1}^m a_{qr} \nonumber \\
    &= \delta\Bigg[\frac{1}{m}\sum_{r=1}^m a_{lr} - \frac{1}{m^2}\sum_{q,r=1}^m a_{qr} \Bigg] \nonumber \\
    &= \delta\Bigg[\frac{1}{m^2}\sum_{q,r=1}^m a_{lr} - \frac{1}{m^2}\sum_{q,r=1}^m a_{qr} \Bigg] \nonumber \\
    &= \frac{\delta}{m^2}\sum_{q,r=1}^m (a_{lr} - a_{qr}),
\end{align}

\noindent which leads directly to condition \eqref{WSAMcond:Hk}. Substituting conditions \eqref{WSAMcond:Lk} and \eqref{WSAMcond:Hk} into the general condition \eqref{WSAMcond:both_short} yields

\begin{equation} \label{WSAMcond:both_long}
   L_l + \mu nH_l = \frac{1}{m}\sum_{q=1}^{m} (a_{ll} + a_{lq} - a_{ql} - a_{qq}) + \frac{\mu n}{m^2}\sum_{q,r=1}^m(a_{lr} - a_{qr}) > 0,
\end{equation}

\noindent a single condition that is able to capture the favorability of strategy $l$ for any mutation rate.

In our framework, we have $d$ distinct possible phases, $\phi_j = j\frac{2\pi}{d}$, arranged symmetrically on the circle, and so there are $m = 2d$ overall strategies: $\{(C, \phi_1), (C, \phi_2), \dots, (C, \phi_d), (N, \phi_1), \linebreak \dots, (N, \phi_d)\}$. Thus, we have a $2d$-by-$2d$ payoff matrix, $A = (a_{qr})$, given by

\begin{center}
\begin{equation}
\resizebox{\columnwidth}{!}{$
    A = \left[\begin{array}{c|c|c|c|c|c|c|c}
    B(0)-c & B(\Delta\phi_{1,2})-c & \dots & B(\Delta\phi_{1,d})-c & \beta(0)-c & \beta(\Delta\phi_{1,2})-c & \dots & \beta(\Delta\phi_{1,d})-c \\
     \hline
     B(\Delta\phi_{1,2})-c & B(0)-c & \dots & B(\Delta\phi_{2,d})-c & \beta(\Delta\phi_{1,2})-c & \beta(0)-c & \dots & \beta(\Delta\phi_{2,d})-c \\
     \hline
     \vdots & \vdots & \ddots & \vdots & \vdots & \vdots & \ddots & \vdots \\
     \hline
     B(\Delta\phi_{1,d})-c & B(\Delta\phi_{2,d})-c & \dots & B(0)-c & \beta(\Delta\phi_{1,d})-c & \beta(\Delta\phi_{2,d})-c & \dots & \beta(0)-c \\
     \hline
     \beta(0) & \beta(\Delta\phi_{1,2}) & \dots & \beta(\Delta\phi_{1,d}) & 0 & 0 & \dots & 0 \\
     \hline
     \beta(\Delta\phi_{1,2}) & \beta(0) & \dots & \beta(\Delta\phi_{2,d}) & 0 & 0 & \dots & 0 \\
     \hline
     \vdots & \vdots & \ddots & \vdots & \vdots & \vdots & \ddots & \vdots \\
     \hline
     \beta(\Delta\phi_{1,d}) & \beta(\Delta\phi_{2,d}) & \dots & \beta(0) & 0 & 0 & \dots & 0 \\
    \end{array}\right].
$}
\end{equation}
\end{center}

\noindent Note that $\Delta\phi_{qr} - \Delta\phi_{rq}$. 

We first use conditions \eqref{WSAMcond:Lk}, \eqref{WSAMcond:Hk}, and \eqref{WSAMcond:both_long} to examine the relative success of $(C,\star)$ strategies in this new framework. Let $l$ denote a $(C,\star)$ strategy. As we have arranged the strategies so that the first $d$ are communicative and the last $d$ are non--communicative, we can split the sums in equations \eqref{WSAMcond:Lk} and \eqref{WSAMcond:Hk} and substitute the appropriate payoffs in each case. Beginning with the low mutation case, we get

\begin{align}
    L_l &= \frac{1}{2d}\Bigg[\sum_{q=1}^{d}(a_{ll} + a_{lq} - a_{ql} - a_{qq}) + \sum_{q=d+1}^{2d} (a_{ll} + a_{lq} - a_{ql} - a_{qq}) \Bigg] \nonumber \\
    &= \frac{1}{2d}\Bigg[\sum_{q=1}^d \Big( (B(0)-c) + (B(\Delta\phi_{ql})-c) - (B(\Delta\phi_{ql})-c) - (B(0)-c) \Big) \nonumber \\
    &\qquad + \sum_{q=d+1}^{2d} \Big( (B(0)-c) + (\beta(\Delta\phi_{q-d,l})-c) - (\beta(\Delta\phi_{q-d,l})) - (0) \Big) \Bigg] \nonumber \\
    &= \frac{1}{2d}\Bigg[\sum_{q=d+1}^{2d} \Big( B(0) - 2c \Big) \Bigg] \nonumber \\
    &= \frac{1}{2}\Big(B(0) - 2c\Big) \nonumber \\
    &= \frac{1}{2}B(0) - c.
\end{align}

\noindent Thus, when the mutation rate is low, a communicative strategy $l$ is preferred when $B(0) > 2c$, confirming our results in Sections \ref{Sec:WeakSel_LowMut} and \ref{Sec:AnySel_LowMut}. 

Similarly, when the mutation rate is high, we apply the same strategy to condition \eqref{WSAMcond:Hk}: 

\begin{footnotesize}
\begin{align}
    H_l &= \frac{1}{4d^2}\Bigg[ \sum_{q=1}^d\sum_{r=1}^d (a_{lr} - a_{qr}) + \sum_{q=1}^{d}\sum_{r=d+1}^{2d} (a_{lr} - a_{qr}) + \sum_{q=d+1}^{2d}\sum_{r=1}^d (a_{lr} - a_{qr}) + \sum_{q=d+1}^{2d}\sum_{r=d+1}^{2d} (a_{lr} - a_{qr}) \Bigg] \nonumber \\
    &= \frac{1}{4d^2}\Bigg[ \sum_{q=1}^d\sum_{r=1}^d \big((B(\Delta\phi_{lr})-c) - (B(\Delta\phi_{qr})-c)\big) + \sum_{q=1}^{d}\sum_{r=d+1}^{2d} \big((\beta(\Delta\phi_{l,r-d})-c) - (\beta(\Delta\phi_{q,r-d})-c)\big) \nonumber \\
    &\qquad + \sum_{q=d+1}^{2d}\sum_{r=1}^d \big((B(\Delta\phi_{lr})-c) - (\beta(\Delta\phi_{q-d,r}))\big) + \sum_{q=d+1}^{2d}\sum_{r=d+1}^{2d} \big((\beta(\Delta\phi_{l,r-d})-c) - (0)\big) \Bigg] \nonumber \\
    &= \frac{1}{4d^2}\Bigg[ -2d^2c + \sum_{q=1}^d\sum_{r=1}^d(B(\Delta\phi_{lr})-B(\Delta\phi_{qr})) + \sum_{q=1}^d\sum_{r=d+1}^{2d}(\beta(\Delta\phi_{l,r-d})-\beta(\Delta\phi_{q,r-d})) \nonumber \\
    &\qquad + \sum_{q=d+1}^{2d}\sum_{r=1}^d (B(\Delta\phi_{lr})-\beta(\Delta\phi_{q-d,r})) + \sum_{q=d+1}^{2d}\sum_{r=d+1}^{2d}(\beta(\Delta\phi_{l,r-d})) \Bigg]. \nonumber \\
    &= \frac{-c}{2} + \frac{1}{4d^2}\Bigg[2d\sum_{r=1}^d B(\Delta\phi_{lr}) - \sum_{q=1}^d\sum_{r=1}^d B(\Delta\phi_{qr}) + 2d\sum_{r=1}^d \beta(\Delta\phi_{lr}) - 2\sum_{q=1}^d\sum_{r=1}^d \beta(\Delta\phi_{qr})\Bigg] \nonumber \\
    &= \frac{-c}{2} + \frac{1}{2d}\sum_{r=1}^d B(\Delta\phi_{lr}) + \frac{1}{2d}\sum_{r=d+1}^{2d} \beta(\Delta\phi_{lr}) - \frac{1}{2d^2}\sum_{q=1}^d\sum_{r=d+1}^{2d} \beta(\Delta\phi_{qr}) - \frac{1}{4d^2}\sum_{q=1}^d\sum_{r=d+1}^{2d} B(\Delta\phi_{qr}) \nonumber \\
    &= \frac{-c}{2} + \frac{1}{2}\langle B(\Delta\phi_{l\cdot})\rangle + \frac{1}{2}\langle \beta(\Delta\phi_{l\cdot})\rangle - \frac{1}{2}\langle\beta(\Delta\phi_{\cdot\cdot})\rangle - \frac{1}{4}\langle B(\Delta\phi_{\cdot\cdot})\rangle \nonumber \\
    &= \frac{1}{2}\Bigg[ \Big(\langle B(\Delta\phi_{l\cdot})\rangle - \frac{1}{2}\langle B(\Delta\phi_{\cdot\cdot})\rangle\Big) + \Big(\langle \beta(\Delta\phi_{l\cdot})\rangle - \langle \beta(\Delta\phi_{\cdot\cdot})\rangle\Big) - c \Bigg].
\end{align}
\end{footnotesize}

\noindent Thus, when the mutation rate is high, a communicative strategy $l$ is favored when 

\begin{equation}\label{WSAM:cond_H_k_long}
    \Big(\langle B(\Delta\phi_{l\cdot})\rangle - \frac{1}{2}\langle B(\Delta\phi_{\cdot\cdot})\rangle\Big) + \Big(\langle\beta(\Delta\phi_{l\cdot})\rangle - \langle\beta(\Delta\phi_{\cdot\cdot})\rangle\Big) > c.
\end{equation}

\noindent Applying condition \eqref{WSAMcond:both_long}, we can conclude that a communicative strategy $l$ is favored for mutation rate $\mu$ if $L_l + \mu nH_l > 0$, or rather,

\begin{equation}
    B(0) + \mu n\Bigg(\Big(\langle B(\Delta\phi_{l\cdot})\rangle - \frac{1}{2}\langle B(\Delta\phi_{\cdot\cdot})\rangle\Big) + \Big(\langle\beta(\Delta\phi_{l\cdot})\rangle - \langle\beta(\Delta\phi_{\cdot\cdot})\rangle\Big)-c\Bigg) > 2c.
\end{equation}

We can derive similar conditions for when a non--communicative strategy $l$ is favored, once again using conditions \eqref{WSAMcond:Lk}, \eqref{WSAMcond:Hk}, and \eqref{WSAMcond:both_long}. When the mutation rate is low, we have 

\begin{align}
    L_l &= \frac{1}{2d}\Bigg[\sum_{q=1}^{d}(a_{ll} + a_{lq} - a_{ql} - a_{qq}) + \sum_{q=d+1}^{2d} (a_{ll} + a_{lq} - a_{ql} - a_{qq}) \Bigg] \nonumber \\
    &= \frac{1}{2d}\Bigg[\sum_{q=1}^d \Big( (0) + (\beta(\Delta\phi_{ql})) - (\beta(\Delta\phi_{ql})-c) - (B(0)-c) \Big) \nonumber \\
    &\qquad + \sum_{q=d+1}^{2d} \Big( (0) + (0) - (0) - (0) \Big) \Bigg] \nonumber \\
    &= \frac{1}{2}\Big(2c - B(0)\Big),
\end{align}

\noindent which gives us the symmetric condition to our earlier results: a non--communicative strategy $l$ is preferred when $B(0) < 2c$.

When the mutation rate is high, we have

\begin{footnotesize}
\begin{align}
    H_l &= \frac{1}{4d^2}\Bigg[ \sum_{q=1}^d\sum_{r=1}^d (a_{lr} - a_{qr}) + \sum_{q=1}^{d}\sum_{r=d+1}^{2d} (a_{lr} - a_{qr}) + \sum_{q=d+1}^{2d}\sum_{r=1}^d (a_{lr} - a_{qr}) + \sum_{q=d+1}^{2d}\sum_{r=d+1}^{2d} (a_{lr} - a_{qr}) \Bigg] \nonumber \\
    &= \frac{1}{4d^2}\Bigg[ \sum_{q=1}^d\sum_{r=1}^d \big((\beta(\Delta\phi_{l-d,r})) - (B(\Delta\phi_{qr})-c)\big) + \sum_{q=1}^{d}\sum_{r=d+1}^{2d} \big((0) - (\beta(\Delta\phi_{q,r-d})-c)\big) \nonumber \\
    &\qquad + \sum_{q=d+1}^{2d}\sum_{r=1}^d \big((\beta(\Delta\phi_{l-d,r})) - (\beta(\Delta\phi_{q-d,r}))\big) + \sum_{q=d+1}^{2d}\sum_{r=d+1}^{2d} \big((0) - (0)\big) \Bigg] \nonumber \\
    &= \frac{c}{2} + \frac{1}{4d^2}\Bigg[2d\sum_{r=1}^d \beta(\Delta\phi_{l-d,r}) - \sum_{q=1}^d\sum_{r=1}^d B(\Delta\phi_{qr}) - 2\sum_{q=1}^d\sum_{r=1}^d \beta(\Delta\phi_{qr}) \Bigg] \nonumber \\
    &= \frac{c}{2} + \frac{1}{2}\langle\beta(\Delta\phi_{l-d,\cdot})\rangle - \frac{1}{4}\langle B(\Delta\phi_{\cdot\cdot}) \rangle - \frac{1}{2} \langle\beta(\Delta\phi_{\cdot\cdot})\rangle \nonumber \\
    &= \frac{1}{2}\Bigg[ c + \langle\beta(\Delta\phi_{l-d,\cdot})\rangle - \langle\beta(\Delta\phi_{\cdot\cdot})\rangle  - \frac{1}{2}\langle B(\Delta\phi_{\cdot\cdot})\rangle \Bigg].
\end{align}
\end{footnotesize}

\noindent Thus, when the mutation rate is high, a non--communicative strategy $l$ is favored when 

\begin{equation}
    \langle B(\Delta\phi_{\cdot\cdot})\rangle + 2\Big(\langle \beta(\Delta\phi_{\cdot\cdot})\rangle - \langle\beta(\Delta\phi_{l-d,\cdot}\rangle \Big) < 2c.
\end{equation}

\noindent Finally, applying condition \eqref{WSAMcond:both_short} allows us to conclude that a non--communicative strategy $l$ is favored without regard to mutation rate if $L_l + \mu nH_l > 0$, or rather, if 

\begin{equation}
    B(0) - \mu n\Big( c + \langle\beta(\Delta\phi_{l-d,\cdot})\rangle - \langle\beta(\Delta\phi_{\cdot\cdot})\rangle  - \frac{1}{2}\langle B(\Delta\phi_{\cdot\cdot})\rangle\Big) < 2c.
\end{equation}

\section{Discussion}
\label{Dis}

Our results describe various conditions under which the synchronization in the SCN of various organisms (i.e. mammals, drosophila, etc. \cite{Sabado2017}) is favored. Of course, not all organisms exhibit behavior that follows a circadian rhythm. For example, some organisms that live in extreme environments (the absence of light, for example), are going through extreme life stages (e.g. migration, reproduction), or are highly social fail to exhibit circadian behavior \cite{moran2014eyeless,bloch2013animal}. It is also the case that not all organisms with circadian rhythms have a circadian system controlled by a ``master clock" like the mammalian SCN. For example, many fish are believed to have a more complex circadian clock arrangement involving a network of interconnected circadian units \cite{moran2014eyeless}. Our results, then, characterize when the SCN is able to function as the ``master clock" to maintain an organism's circadian system. 

Taken together, the various conditions under which $(C,\star)$ strategies are favored under each set of assumptions above begin to give us a picture of what characteristics are necessary for this kind of synchronization behavior to occur in the neurons of such organisms' SCN. Broadly, we found that the benefits associated to communication must sufficiently outweigh the cost for communication to be favored and synchronization to occur. One specific, ubiquitous example of this theme is our finding that the largest possible benefit, $B(0)$, must exceed twice the cost of communication, $c$. In Section \ref{Sec:WeakSel_LowMut}, we found that under the assumptions of weak selection, $B(0) > 2c$ is necessary for communicative strategies to be favored by the pairwise invasion dynamics, as seen in condition \eqref{WSLMcond:rhoE_2_rhoF}. This is echoed again in Section \ref{Sec:StrSel_LowMut}, where we found, by combining conditions \eqref{SSLMeqn:cond1} and \eqref{SSLMeqn:cond2}, that communication will always be favored when $c < \beta(\Delta\phi) < B(0) - c$, or equivalently, when $2c < \beta(\Delta\phi) + c < B(0)$. Thus, we again see here the need for $B(0)$ to exceed twice the cost of communication. We proved this condition rigorously, using a different characterization of favorability, in Section \ref{Sec:AnySel_LowMut}. It appears again in the low mutation case (condition \eqref{WSAMcond:Lk}) of Section \ref{Sec:WeakSel_AnyMut}. The ubiquity of this condition in our results suggests that this is an important characterization of those organisms who exhibit this kind of synchronization behavior in their SCNs. 

A closer examination of the other results obtained above furthers the broad idea that the benefits associated to communication must sufficiently exceed the cost for communication to be favored and synchronization to occur. For example, in the case of strong selection and low mutation, conditions \eqref{SSLMeqn:cond1} and \eqref{SSLMeqn:cond2} demonstrate both that $\beta(\Delta\phi)$ must exceed the cost of communication, but also that $B(0)$ must sufficiently exceed $\beta(\Delta\phi)$. This relationship ($c < \beta(\Delta\phi) < B(0) - c$) also implies condition \eqref{WSLMcond:neut} in Section \ref{Sec:WeakSel_LowMut}, which describes when $(C,\star)$ strategies perform better than strategies under the neutral process. 

This idea can also be seen in the high mutation case of Section \ref{Sec:WeakSel_AnyMut}. In condition \eqref{WSAMcond:both_long}, when the mutation probability $\mu$ is sufficiently large, the condition characterizing the favorability of a communicative strategy $l$ simplifies to require 

\begin{equation}
    \Big(\langle B(\Delta\phi_{l\cdot})\rangle - \frac{1}{2}\langle B(\Delta\phi_{\cdot\cdot})\rangle\Big) + \Big(\langle\beta(\Delta\phi_{l\cdot})\rangle - \langle\beta(\Delta\phi_{\cdot\cdot})\rangle\Big)-c > 0.
\end{equation}

\noindent The first term here describes the difference between the average $B$ benefit for strategy $l$ with the average $B$ benefit of all other strategies; the second term does the same for the average $\beta$ benefit of strategy $l$. Thus, this condition requires that the sum of the average extra benefit a neuron is awarded for having the strategy $l$ must exceed the cost of communication in order for the communicative strategy $l$ to be favored.

\subsection{Conclusions}

These initial explorations have affirmed the intuitive idea that the observed synchronization in the SCN requires that the benefits of communication sufficiently outweigh the cost: the largest possible benefit a neuron may receive by communicating must exceed twice the cost incurred by communicating.  This result is robust as it holds across multiple different assumptions about the system and represents a foundational step in undrestanding the impact of evolutionary constraints and trade-offs in the development of synchronizable circadian systems.   


Topological properties of networks of coupled oscillators play a critical role in determining whether such a system will synchronize. Differences in topology can promote strong synchronization or weaker partial synchronization in a dizzying array of patterns - waves \cite{hong2011kuramoto}, chimeric states \cite{abrams2005chimera}, cluster synchronization \cite{allefeld2007eigenvalue,favaretto2017cluster,ji2013cluster,pecora2014cluster,zhou2006hierarchical}, pinwheels \cite{abrams2005chimera}, and combinations of these. On the other hand, researchers in evolutionary game theory have explored the impact of topology on the emergence of cooperation among agents in a structured population \cite{nowak2010evolutionary}. Here again, we see a variety of outcomes -- systems that converge to complete cooperation, complete defection, or mixed populations of defectors and cooperators -- and a large body of work delineates topological structures that facilitate cooperation~\cite{nowak2010evolutionary,lieberman2005evolutionary,ohtsuki2006simple,santos2005scale,szabo2007evolutionary}. A recent sequence of papers describe topological statistics and signatures that push a system towards cooperation in pairwise interactions~\cite{antal2009evolution,tarnita2009evolutionary,allen2017evolutionary} and high-order interactions~\cite{alvarez2020evolutionary}. This work provides a first step in exploring how these topologies arise in the context of evolutionary processes.  While this initial work applies only to the simplest case of developing a fully connected system of oscillators, the framework easily adapts to more flexible methods of building connection topologies.

\section*{Author contributions}
E.A.T., F.F., \& S.D.P. conceived the model and performed theoretical analysis; E.A.T. \& S.D.P. plotted figures, and wrote the first version of the draft; S.D.P. secured funding and supervised the project; E.A.T., F.F., S.D.P. contributed to the revision of the draft and gave approval of the final manuscript.

\section*{Acknowledgments}
This work is supported by NSF IOS-1730508. F.F. is grateful for the generous financial support by the Bill \& Melinda Gates Foundation (award no. OPP1217336), the NIH COBRE Program (grant no. 1P20GM130454), and the Neukom CompX Faculty Grant.



\end{document}